\documentclass[12pt]{iopart}
\baselinestretch
\usepackage{amssymb}
\usepackage[dvips]{graphicx}
\usepackage{colordvi}
\usepackage{times}

\newcommand{\un}[1]{\underline{#1}}

\newcommand{\bc}{\begin{center}}
\newcommand{\ec}{\end{center}}
\newcommand{\be}{\begin{equation}}
\newcommand{\ee}{\end{equation}}
\newcommand{\bna}{\begin{eqnarray}}
\newcommand{\ena}{\end{eqnarray}}

\newcommand{\mpaa}{\begin{minipage}[t]{6cm}}
\newcommand{\mpea}{\end{minipage}}
\newcommand{\mpab}{\begin{minipage}[t]{8cm}}
\newcommand{\mpeb}{\end{minipage}}
\newcommand{\mpac}{\begin{minipage}[t]{13cm}}
\newcommand{\mpec}{\end{minipage}}
\newcommand{\mpad}{\begin{minipage}[t]{13cm}}
\newcommand{\mped}{\end{minipage}}
\newcommand{\mpae}{\begin{minipage}[t]{13cm}}
\newcommand{\mpee}{\end{minipage}}
\newcommand{\mpaf}{\begin{minipage}[t]{6cm}}
\newcommand{\mpef}{\end{minipage}}

\newcommand{\zd}{\delta}
\newcommand{\ze}{\epsilon}

\newcommand {\bdm} {\begin{displaymath}}
\newcommand {\edm} {\end{displaymath}}

\usepackage{color}
\definecolor{darkblue}{rgb}{0,0,0.6}
\definecolor{darkred}{rgb}{0.7,0,0}
\definecolor{darkgreen}{rgb}{0,0.7,0}
\usepackage{amsfonts}

\newcommand{\bea}{\begin{eqnarray}}
\newcommand{\eea}{\end{eqnarray}}

\begin{document}

\title{Fluctuation-dissipation relation for chaotic non-Hamiltonian systems}

\author{Matteo Colangeli$^1$, Lamberto Rondoni$^{1,2}$, Angelo Vulpiani $^{3,4}$}
\address{$^1$Dipartimento di Matematica, Politecnico di Torino, Corso Duca degli Abruzzi 24, 10129 Torino, Italy\\
$^2$INFN, Sezione di Torino, Via P. Giuria 1, 10125 Torino, Italy \\
$^3$Dipartimento di Fisica, Universit\`a di Roma Sapienza, p.le Aldo Moro 2, 00185 Roma, Italy \\
$^4$Istituto dei Sistemi Complessi (ISC-CNR), Via dei Taurini 19, 00185 Roma, Italy}

\ead{colangeli@calvino.polito.it}

\begin{abstract}
In dissipative dynamical systems phase space volumes contract, on average. Therefore, 
the invariant measure on the attractor is singular with respect to the Lebesgue measure. 
As noted by Ruelle, a generic perturbation pushes the state out of the attractor, hence
the statistical features of the perturbation and, in particular, of the relaxation, cannot 
be understood solely in terms of the unperturbed dynamics on the attractor. This remark 
seems to seriously limit the applicability of the standard fluctuation dissipation procedure
in the statistical mechanics of nonequilibrium (dissipative) systems.
In this paper we show that the singular character of the steady state does not constitute a serious limitation in
the case of systems with many degrees of freedom. The reason is that one typically deals with
projected dynamics, and these are associated with regular probability distributions in the
corresponding lower dimensional spaces.
\end{abstract}

\maketitle

\section{Introduction}

Since its early developments, due mainly to the works of L.\ Onsager and R.\ Kubo \cite{Onsager,GK,G1,Kubo}, 
the fluctuation-dissipation theorem (FDT) represents a cornerstone in the construction of a 
theory of nonequilibrium phenomena \cite{VulpiDyn}. This celebrated result was developed in 
the context of Hamiltonian dynamical systems, slightly perturbed out of their thermodynamic 
equilibrium, and it was later extended to stochastic systems obeying a Langevin Equation \cite{BPRV,CMW}. 
The importance of the FDT rests on the fact that it sheds light on the crucial 
relation between the response $R_V(t)$ of a system to an external perturbation and a time correlation
function computed at \textit{equilibrium}. In other words, having perturbed a 
given Hamiltonian $H_0$ with an external field $h_e$, to obtain the perturbed Hamiltonian 
$H_0 -h_e V$, where $V$ is an observable conjugated with $h_e$, the FDT allows us to 
compute nonequilibrium quantities, such as the transport coefficients \cite{Zwanzig,Matt4}, 
solely in terms of the unperturbed equilibrium state.
On the other hand, a generic dynamical system is not Hamiltonian: for phenomenological 
practical pruposes, one typically deals with dissipative dynamics, as in the important 
case of viscous hydrodynamics \cite{BPRV}.

The invariant measure of a chaotic dissipative system, $\mu$ say, is singular with respect to 
the Lebesgue measure and is usually supported on a fractal attractor. This is not just a
mathematical curiosity, it is 
a potential source of difficulties for the applicability of the FDT in dissipative systems. 
Indeed, the standard FDT ensures that the statistical features of a perturbation are 
related to the statistical properties of the unperturbed system, but that cannot 
be the case in general, in dissipative systems. The reason is that, given an initial 
state $\un{x}(0)$ on the 
attractor and a generic perturbation $\delta \un{x}(0)$, the perturbed initial state 
$\un{x}^{(p)}(0)=\un{x}(0)+\delta \un{x}(0)$ and its time evolution may lie outside 
the support of $\mu$, hence their statistical properties cannot be
expressed by $\mu$, which attributes vanishing probability to such states.
In the cases considered by Ruelle \cite{Ruelle}, the perturbation $\delta \un{x}(0)$ and its 
time evolution $\delta \un{x}(t)$ can be decomposed as the sum of two parts, 
$\delta \un{x}_{\perp}(t)$ and $\delta \un{x}_{\parallel}(t)$, respectively 
perpendicular and parallel to the ``fibres'' of the attractor, 
$$
\delta \un{x}(t)=\delta \un{x}_{\perp}(t)+\delta \un{x}_{\parallel}(t)
$$
which makes it natural to expect that the statistical features of $\delta \un{x}_{\parallel}(t)$ 
be related to the dynamics on the attractor, while it is easy to construct examples in which
$\delta \un{x}_{\perp}(t)$ is not described by the unperturbed dynamics.

From the mathematical point of view, this fact is rather transparent. On the other hand,
it should not be a concern in statistical mechanics, except in pathological cases.
Indeed, a series of numerical investigations of 
chaotic dissipative systems shows that the standard FDT holds 
under rather general conditions, mainly if the invariant measure is absolutely continuous with respect to Lebesgue, cf.\ Ref.\cite{BPRV} for a review. 
Moreover, although dissipative systems have singular invariant measures, any small amount of 
noise produces smooth invariant measures, which allow generalized FDTs to be expressed solely 
in terms of the unperturbed states, analogously to the standard equilibrium case. Apart from  
technical aspects, the intuitive reason 
for which the FDT in systems with noise can be expressed only in terms of the invariant 
measure, is that $\un{x}^{p}(0)$ remains within the support of this measure. 

In this paper, we want to take advantage of the fact that a similar situation is realized 
without any noise, if one works in the projected space of the physically relevant observables. 
Indeed, marginals of singular phase space measures, on spaces of sufficiently lower dimension 
than the phase space, are usually regular \cite{EvRon,BKL2005}.

Our paper is organized as follows: Section \ref{sec:sec2} is devoted to a short presentation of some general results on FDT for chaotic dissipative systems. In Sec.\ \ref{sec:sec3} we discuss the numerical results for two dissipative chaotic maps, showing that the singular character of their 
invariant measures does not prevent the response of standard observables to be expressed only 
in terms of the invariant measure, as in the standard case.
Conclusions are drawn in Sec.\ \ref{sec:sec4}.

\section{Some results on FDT in chaotic dissipative systems}
\label{sec:sec2}
\begin{figure}
\centering
\fbox{\includegraphics[width=5.5cm]{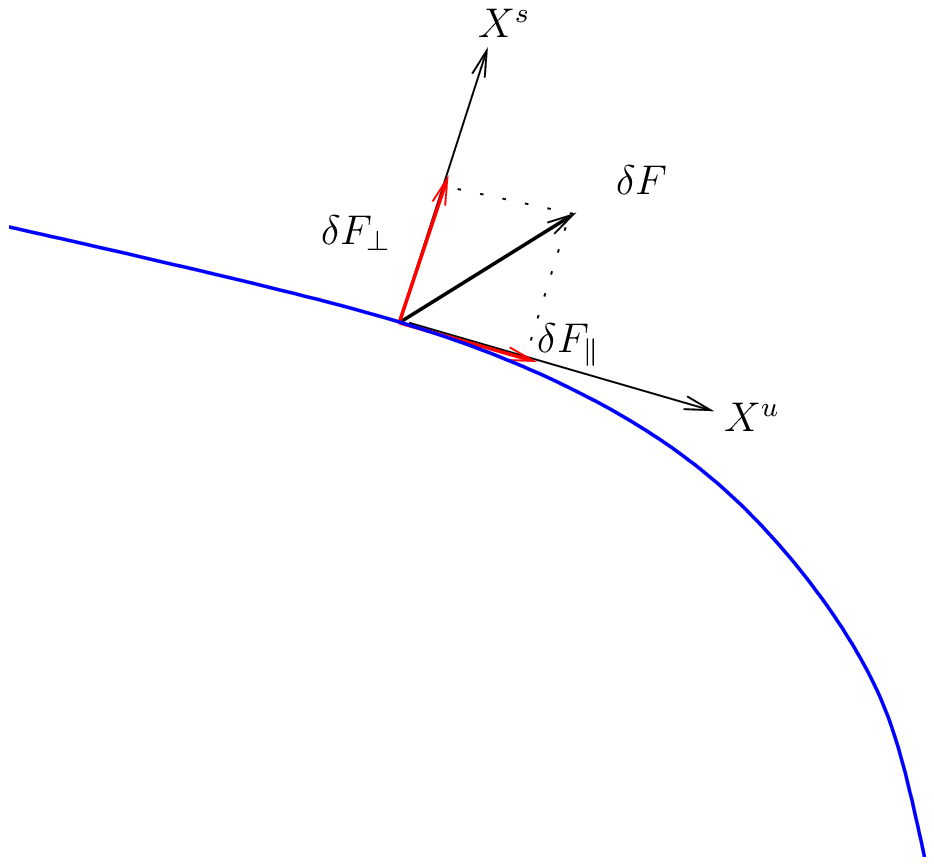}}
\hspace{2mm}
\fbox{\includegraphics[width=5.5cm]{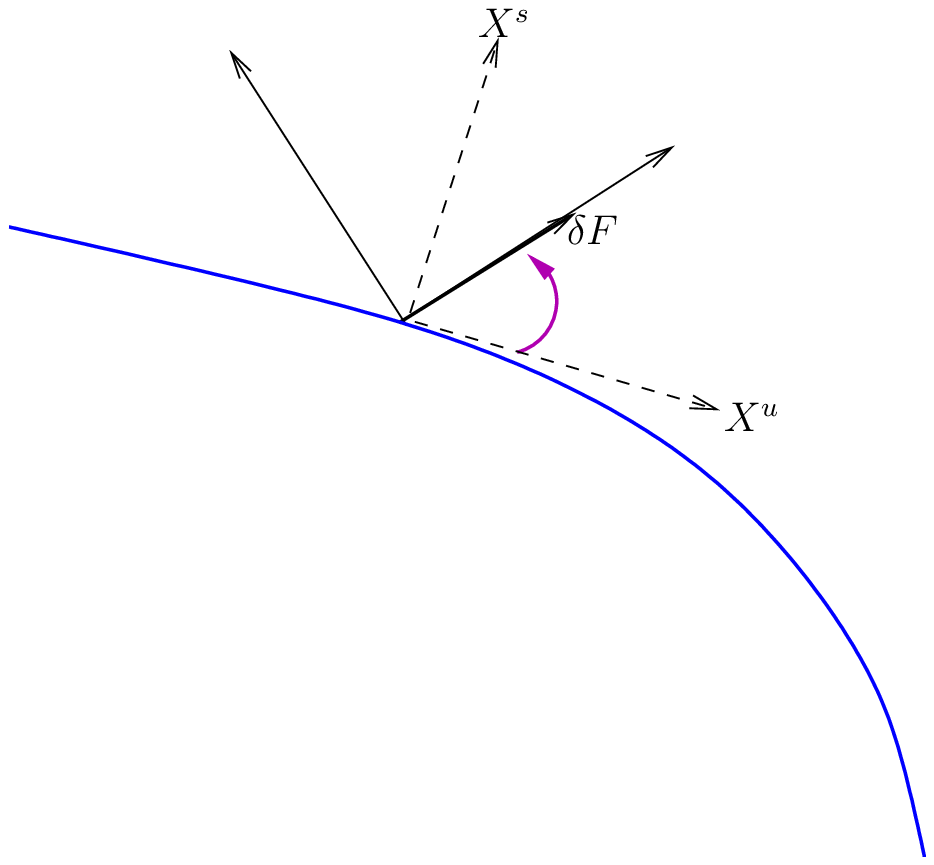}}
\caption{\textit{Left panel}: In Ruelle's approach, the perturbation is expressed as the sum of
one component parallel to the unstable manifold and one parallel to the stable manifold.
\textit{Right panel}: In the present work, the reference frame is rotated so that the direction
of the perturbation coincides with one of the basis vectors.}
\label{compar}
\end{figure}

Let us concisely recall Ruelle's approach to linear response in deterministic dissipative 
dynamical systems \cite{Ruelle}.
Let ($\mathcal{M},S^t,\mu$) be a dynamical system, with $\mathcal{M}$ its compact phase space, $S^t:\mathcal{M}\rightarrow \mathcal{M}$ a one parameter group
of diffeomorphisms and $\mu$ the invariant natural measure. 
Following Ruelle \cite{Ruelle}, who considers axiom A systems, one may show that the effect of a perturbation 
$\delta F(t)=\delta F_{\parallel}(t)+\delta F_{\perp}(t)$ on the response of a generic 
(smooth enough) observable $A$ attains the form:
\be
\delta \overline{A}(t)=\int_{0}^{t} R_{\parallel}^{(A)}(t-\tau) \delta F_{\parallel}(\tau) d\tau+\int_{0}^{t} R_{\perp}^{(A)}(t-\tau) \delta F_{\perp}(\tau) d\tau \label{4}
\ee
where the subscript $_\parallel$ refers to the dynamics on the unstable tangent bundle (along 
the attractor), while $_\perp$ refers to the transversal directions, cf.\ left panel of
Fig.\ \ref{compar}. 
Ruelle's central remark is that $R_{\parallel}^{(A)}$ may be expressed in terms of 
a correlation function evaluated with respect to the unperturbed dynamics, while 
$R_{\perp}^{(A)}$ depends on the dynamics along the stable manifold, hence it may not 
be determined by $\mu$, and should 
be quite difficult to compute numerically \cite{BPRV}.

To illustrate these facts, the Authors of Ref.\cite{Sepul} study a $2$-dimensional model, 
which consists of a chaotic rotator on a plane and, for such a system, succeed to numerically estimate the  $R_{\perp}^{(A)}$ term in eq.(\ref{4}). Nevertheless, in the next Section,
we argue that $R_{\perp}^{(A)}$ may spoil the generalized FDT only if the perturbation is 
carefully oriented with respect to the stable and unstable manifolds.
This is only possible in peculiar situations, such as those of Ref.\cite{Sepul}, in which the
invariant measure is the product of a radial and and angular component and, furthermore, the perturbation lies on the radial direction, leaving the angular dynamics unaffected.

A different approach to the FDT has been proposed in \cite{BLMV}, which concerns 
deterministic dynamics perturbed by stochastic contributions. Here, the invariant measure
$\mu$ can be assumed to have density $\rho$: ${\rm d} \mu(\un{x})=\rho(\un{x}) d\un{x}$. Then, 
if the initial 
conditions are modified by an \textit{impulsive} perturbation 
$\un{x}_0\rightarrow \un{x}_0+\delta \un{x}_0$, the invariant density $\rho(\un{x}_0)$ is
replaced by a perturbed initial density 
$\rho_0(\un{x}_0;\delta \un{x}_0)=\rho(\un{x}_0-\delta \un{x}_0)$, where the subscript 0
denotes the initial state, right after the perturbation. This state is not stationary and
evolves in time, producing time dependent densities $\rho_t(\un{x}_0;\delta \un{x}_0)$, which 
are assumed to eventually relax back to $\rho(\un{x}_0)$. 
Given the transition probability $W(\un{x}_0,0\rightarrow \un{x},t)$ determined 
by the dynamics, the response of coordinate $x_i$ is expressed by:
\be
\overline{\delta x_i (t)} = \int \int  x_i 
\left[ \rho(\un{x}_0-\delta \un{x}_0)-\rho(\un{x}_0) \right] 
W(\un{x}_0,0\rightarrow \un{x},t) d\un{x}_0 d\un{x} 
\label{V0}
\ee
and one may introduce the response function $R_{ij}$ as \cite{BLMV}:
\be
R_{ij}(t)=\frac{\overline{\delta x_i (t)}}{\delta x_j(0)}=
-\left\langle x_i (t) \left. \frac{\partial \log \rho}{\partial x_j}\right|_{t=0}\right\rangle \label{V1} 
\ee
which is a correlation function computed with respect to the unperturbed state.
It is worth to note that it makes no difference 
in the derivation of eq.(\ref{V1}) whether the steady state is an equilibrium state 
or not; it suffices that $\rho$ be differentiable. 

Let us consider again Eq.(\ref{V0}) and, for sake of simplicity, assume that 
all components of $\delta \un{x}(0)$ vanish, except the $i$-th component. Then, 
the response of $x_i$ may also be written as:
\bea
\overline{\delta x_i} (t) &=& \int x_i  \left\{\int \left[
\rho(\un{x}_0-\delta \un{x}_0)-\rho(\un{x}_0)\right] W(\un{x}_0,0\rightarrow \un{x},t) d\un{x}_0 \prod_{j\neq i}dx_j\right\} dx_i \nonumber \\
&\equiv& \int x_i B_i(x_i,\delta \un{x}_0,t) dx_i \label{V2}
\eea
where $B_i(x_i,\delta \un{x}_0,t)$, defined by the term within curly brackets,
may also be written as:
\be
B_i(x_i,\delta \un{x}_0,t)
=\widetilde{\rho_t}(x_i;\delta \un{x}_0)-\widetilde{\rho}(x_i) \label{V3}
\ee
where $\widetilde{\rho}(x_i)$ and $\widetilde{\rho_t}$ are the marginal probability 
distributions defined by:
$$
\widetilde{\rho}(x_i)=\int \rho(\un{x}) \prod_{j\neq i}dx_j ~,
\qquad
\widetilde{\rho_t}(x_i;\delta \un{x}_0)=\int \rho_t(\un{x};\delta \un{x}_0) \prod_{j\neq i}dx_j ~.
$$

As projected singular measures are expected to be smooth, especially if the dimension of
the projected space is sensibly smaller than that of the original space, one may adopt the
same procedure also for dissipative deterministic dynamical systems. Indeed, the response
function $B_i(x_i,\delta \un{x}_0,t)$ in Eq.(\ref{V3}) is also expected to be smooth,
and to make the response of $x_i$ computable from the invariant measure only.
In the next section we investigate this possibility.

\section{Coarse graining analysis}
\label{sec:sec3}

In terms of phase space probability measures, the response formula Eq.(\ref{V0}) reads:
\be
\overline{\delta x_i}(t) = 
\int x_i ~ {\rm d} \mu_t(\un{x};\delta \un{x}_0) - \int x_i ~ {\rm d} \mu(\un{x}) 
\label{meas1}
\ee
where ${\rm d} \mu_t(\un{x};\delta \un{x}_0)$ is the time evolving perturbed measure 
whose initial state is given by
$$
{\rm d} \mu_0(\un{x}_0;\delta \un{x}_0)=\rho_0(\un{x}_0;\delta \un{x}_0)~{\rm d} \un{x}_0
=\rho(\un{x}_0-\delta\un{x}_0)~{\rm d} \un{x}_0 ~.
$$
Because dissipative dynamical systems do not have an invariant probability density, it is 
convenient to introduce a coarse graining in phase space, to approximate the singular 
invariant measure $\mu$ by means of piecewise constant distributions. 

Let us consider a $d$-dimensional phase space $\mathcal{M}$, 
with an $\epsilon$-partition made of
a finite set of $d$-dimensional hypercubes $\Lambda_k(\epsilon)$ of side $\epsilon$ and 
centers $\un{x}_k$. Introduce the $\epsilon$-coarse graining of $\mu$ and of $\mu_t$ 
defined by the probabilities $P_k(\epsilon)$ and $P_{t,k}(\epsilon;\delta \un{x}_0)$
of the hypercubes $\Lambda_k(\epsilon)$:
\be 
P_k(\epsilon)=\int_{\Lambda_k(\epsilon)} {\rm d} \mu(\un{x}) \hskip 3pt, \quad 
P_{t,k}(\epsilon;\delta \un{x}_0)
= \int_{\Lambda_k(\epsilon)} {\rm d} \mu_t(\un{x};\delta \un{x}_0) ~.
\label{CG}
\ee
This leads to the coarse grained invariant density $\rho(\un{x};\epsilon)$:
\be
\rho(\un{x};\epsilon) =
\sum_k \rho_k(\un{x};\epsilon) \hskip 3pt, \hskip 6pt \hbox{with} \hskip 6pt 
\rho_k(\un{x};\epsilon) = \left\{
\begin{array}{ll}
    P_k(\epsilon)/\epsilon^d & \hbox{if $ x \in \Lambda_k(\epsilon)$} \\
    0 & \hbox{else}\label{densCG}
  \end{array}\right. 
\ee
Let $Z_i$ be the number of bins of of form 
$\left[x_i^{(q)}-\epsilon/2,x_i^{(q)}+\epsilon/2\right)$, $q\in\{1,2,...,Z_i\}$, 
in the $i$-th direction. Then, the marginalization of the coarse grained distribution 
yields the following set of $Z_i$ probabilities:
\be
\hskip -45pt
p_i^{(q)}(\epsilon) = 
\int_{x_i^{(q)}-\frac{\epsilon}{2}}^{x_i^{(q)}+\frac{\epsilon}{2}} 
\left\{\int \rho(\un{x};\epsilon) \prod_{j\neq i} dx_j \right\}  {\rm d} x_i = 
{\rm Prob}\left(x_i\in\left[x_i^{(q)}-\frac{\epsilon}{2},x_i^{(q)}+
\frac{\epsilon}{2}\right)\right) 
\label{CG2}
\ee
each of which is the invariant probability that the coordinate $x_i$ lie in one of
the $Z_i$ bins.
In an analogous way, one may define the marginal of the evolving coarse grained perturbed
probability $p_{t,i}^{(q)}(\epsilon;\delta \un{x}_0)$. In both cases,
dividing by $\ze$, one obtains the coarse grained marginal probability densities $\rho_i^{(q)}(\epsilon)$ and $\rho_{t,i}^{(q)}(\epsilon;\delta \un{x}_0)$, as well as the $\epsilon$-coarse grained version of the response function $B_i(x_i,\delta \un{x}_0,t)$:
\be
B_{i}^{(q)}(x_i,\delta \un{x}_0,t,\epsilon)
= \frac{1}{\epsilon} \left[p_{t,i}^{(q)}(\epsilon,\delta \un{x}_0)-p_i^{(q)}(\epsilon)\right]
= {\rho_{t,i}^{(q)}(\epsilon,\delta \un{x}_0)-\rho_i^{(q)}(\epsilon)} \label{CG4} 
\ee
In the following, we will show that the r.h.s.\ of Eq.(\ref{CG4}) tends to a regular
function of $x_i$ in the
$Z_i \rightarrow \infty$, $\epsilon \to 0$, limit.
Then, in the limit of small perturbations $\delta \un{x}_0$, 
$B_{i}^{(q)}(x_i,\delta \un{x}_0,t,\epsilon)$ may be expanded as a Taylor series, to yield
an expression similar to standard response theory, in the sense that it depends solely on 
the unperturbed state. The difference, here, is that 
the invariant measure is singular and represents a nonequilibrium steady state.

To illustrate this fact, we run a set of $N$ trajectories with uniformly distributed 
initial conditions in the phase spaces of two simple, but substantially different, 
2-dimensional maps: a dissipative baker map, and the Henon map.

\subsection{The dissipative baker map}
\begin{figure}
   \centering
   \includegraphics[width=8.5cm]{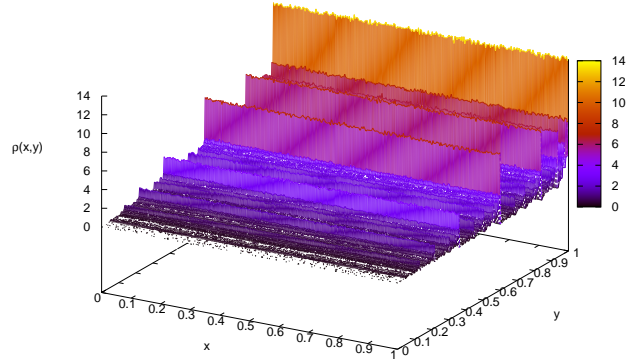}
   \caption{Invariant probability distribution of the map defined by Eq.\ (\ref{Map1}).}
\label{Histo-Baker}
\end{figure}

Let $\mathcal{M} = [0,1] \times [0,1]$ be the phase space, and
consider the evolution equation
\be
\left(\begin{array}{c}
  x_{n+1} \\
  y_{n+1}
\end{array}\right)
=M
\left(\begin{array}{c}
  x_{n} \\
  y_{n}
  \end{array}\right)
  =\left\{
     \begin{array}{ll}
       \left(\begin{array}{c}
                   x_{n}/l \\
                   r y_{n}
                 \end{array}\right), & \hbox{for $0\leq x_n < l$;} \\
       \left(\begin{array}{c}
                   (x_{n}-l)/r \\
                   r+l y_{n}
                 \end{array}\right), & \hbox{for $l\leq x_n \leq 1$.}
     \end{array} \right. \label{Map1} \quad .
\ee
whose Jacobian determinant is given by
\be
J_M(\un{x})=\left\{
  \begin{array}{ll}
    J_A= r/l ~, & \hbox{for $\:0\leq x \leq l$;} \\
    J_B= l/r=J_A^{-1} ~, & \hbox{for $\:l\leq x \leq 1$.}\label{JDorf}
  \end{array}\right. \quad .
\ee
and shows that the $M$ is dissiaptive for $l< 1/2$.
The map $M$ is hyperbolic, since stable and unstable manifolds which intersect each other orthogonally are defined at all points $\un{x}\in \mathcal{M}$, except in the irrelevant 
vertical segment at $x=l$. The directions of these manifolds coincide, respectively, with 
the vertical and horizontal directions.
It can also be shown that this dynamical system is endowed with an invariant measure $\mu$
which is smooth along the unstable manifold and singular along the stable one, cf.\ Figs.\ref{Histo-Baker}. In particular, $\mu$ factorizes as 
${\rm d} \mu(\un{x}) = {\rm d}{x} \times {\rm d} \lambda (y)$, similarly to 
the case of \cite{Sepul}.

\begin{figure}
   \centering
   \includegraphics[width=6.5cm]{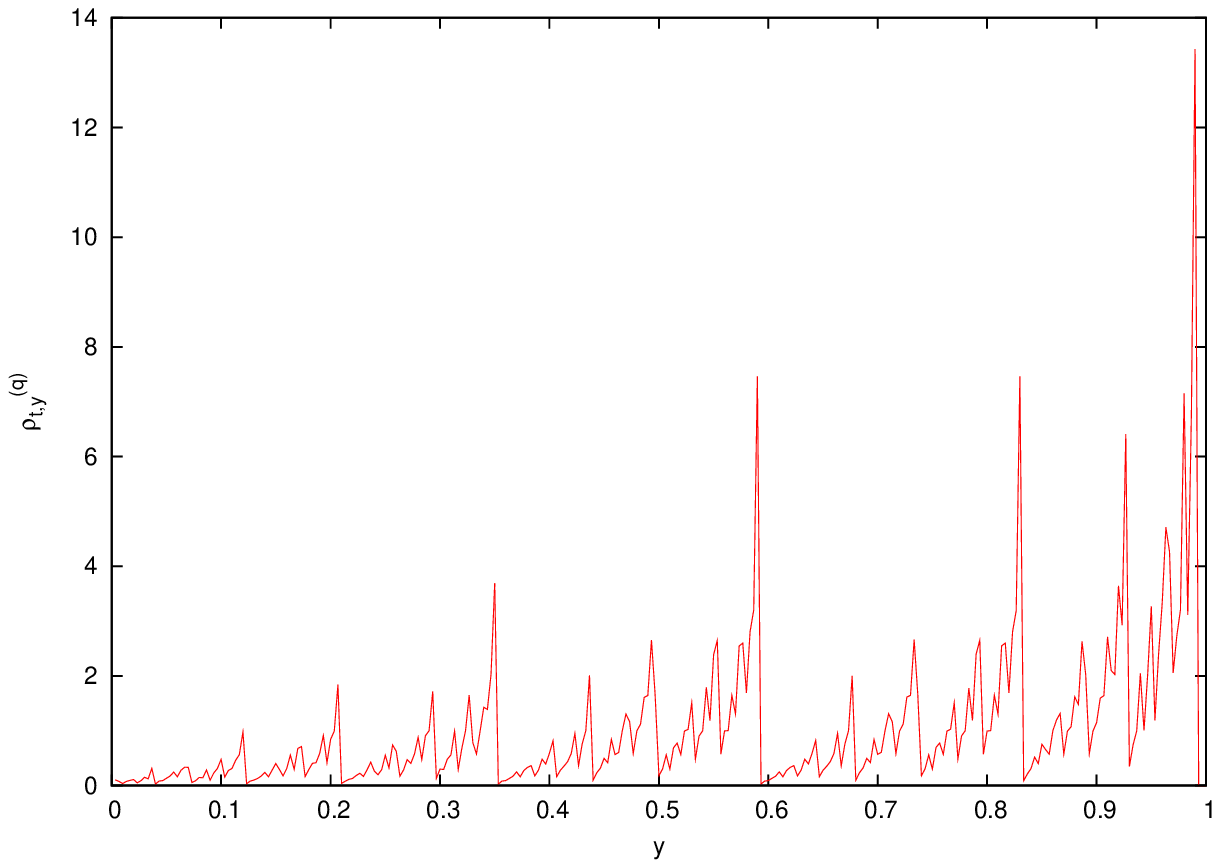}
 \ \hspace{2mm} \
   \includegraphics[width=6.5cm]{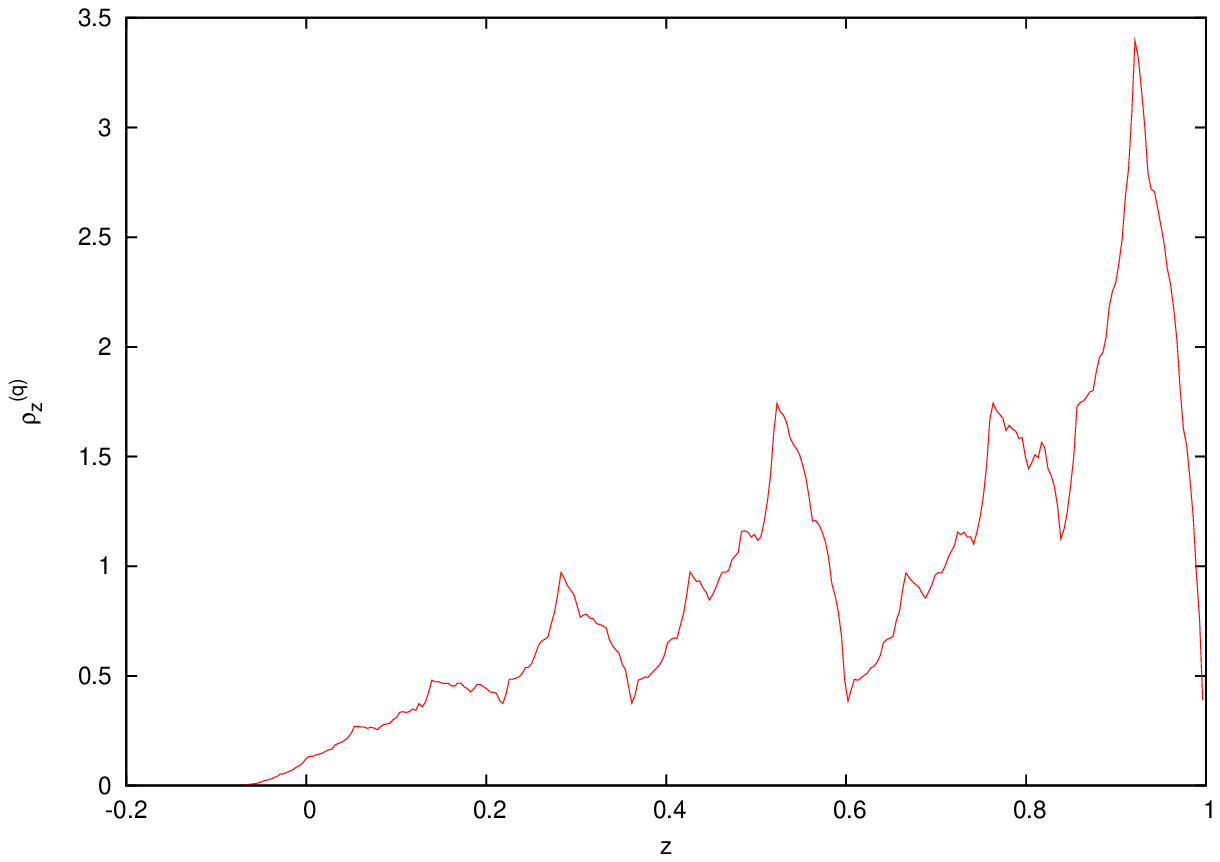}
   \caption{\textit{Left panel}: singular distribution of the unperturbed system, obtained projecting the invariant measure onto the 
vertical direction $y$, which is the direction of the stable manifold. \textit{Right panel}: projected invariant density along the direction of the perturbation, which forms an angle $\alpha=\pi/8$ radiants with the $y$-direction. 
$z=1-(x \cos(\alpha)-y\sin(\alpha)+\sin(\alpha))$
is the coordinate along this direction. Although hardly differentiable, this projected
distribution has a density.}\label{Baker}
\end{figure}

In order to verify whether the functions corresponding to the above introduced $B_{i}^{(q)}(x_i,\delta \un{x}_0,t,\epsilon)$ become regular functions in the 
fine graining limit, let us consider first an 
impulsive perturbation, directed purely along the stable manifold, i.e. 
$\delta\un{x}_0=(0,\delta y_0)$.
Ruelle's work on singular measures is clearly relevant, in this case, because the support 
of the marginal perturbed probability measure, obtained projecting out the $y$-direction 
has simply drifted preserving its singular character, while the state may have fallen
outside the support of the unperturbed invariant measure, cf.\ left panel of 
Fig.\ref{Baker}. 

Consider now an initial impulsive perturbation with one component, no matter how small, 
along the unstable manifold, $\delta\un{x}_0=(\zd x_0,\delta y_0)$ and rotate the vectors of the basis of the 2-dimensional
plane, so that the coordinate $x$ lies along the direction of the perturbation, as shown in the right panel of Fig.\ref{compar}.
We find that $B_{x}^{(q)}(x,\delta \un{x}_0,t,\epsilon)$ is regular as a function of $x$. 
Indeed, the projections of $\mu$ and of its perturbations onto the direction of 
$\delta\un{x}_0$ have a density along all directions except the vertical one, cf.\ right panel of 
Fig.\ref{Baker}. Hence, a small perturbation does not take the state outside the corresponding projected support. 

As already noted in \cite{Sepul}, this Baker map shows that the response to very carefully 
selected perturbations, cannot be computed in general from solely the invariant measure.
However, similarly to the case of \cite{Sepul}, the factorization of $\mu$ makes the present
case rather peculiar. Indeed, for the overwhelming majority of dynamical systems, it looks
impossible to select directions such that the projected measures preserve the same degree
of singularity as the full measures. This is a consequence of the fact that stable and 
unstable manifolds have different orientations in different parts of the phase space, 
provided they exist. Clearly, the 
higher the dimensionality of the phase space and the larger the number of projected out
dimensions, the more difficult it is to preserve singular characters.

\subsection{The Henon map}

Consider for instance the Henon map defined by:
\be
\left(\begin{array}{c}
  x_{n+1} \\
  y_{n+1}
\end{array}\right)
=M
\left(\begin{array}{c}
  x_{n} \\
  y_{n}
  \end{array}\right)
  =\left(
     \begin{array}{c}
                   y_{n}+1-a x_{n}^2 \\
                   b x_{n}
                 \end{array}\right). \label{Map2} \quad 
\ee
one the phase space $\mathcal{M}=[-\frac{3}{2},\frac{3}{2}]\times[\frac{1}{2},\frac{1}{2}]$,
where $a=1.4$ and $b=0.3$ imply a chaotic dissipative dynamics, with a fractal invariant measure
$\mu$, which is not the product of the marginal measures obtained by projecting onto the 
horizontal and the vertical directions. These marginals are indeed regular and would
yield a regular product. 
As stable and unstable manifolds wind around, changing orientation, in a very complicated fashion,
it seems impossible, here, to disentangle the contributions of one phase space direction from the other.
\begin{figure}
   \centering
   \includegraphics[width=6.5cm]{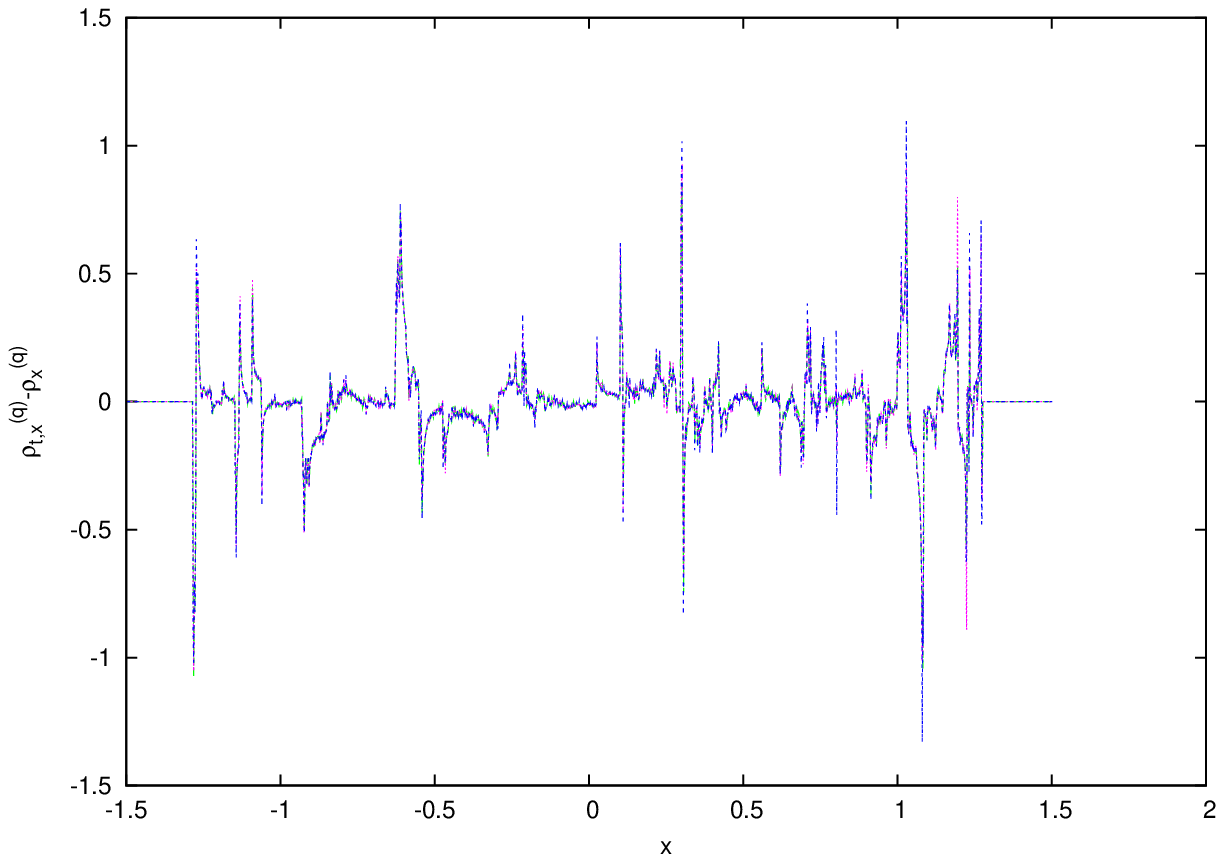}
  \hspace{3mm} \
   \includegraphics[width=6.5cm]{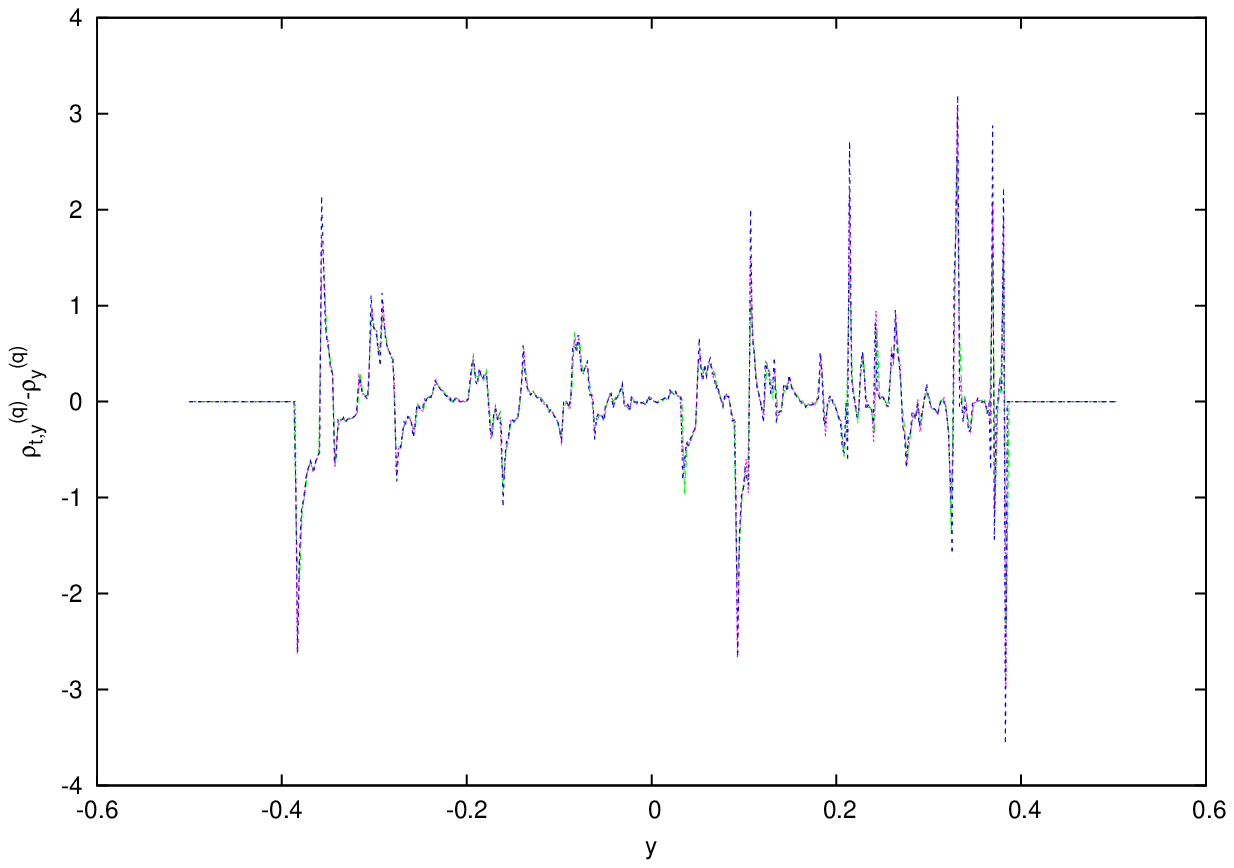}
   \caption{\textit{Left panel}: plot of the function $B_x^{(q)}(x,\delta \un{x}_0,t,\epsilon)$, introduced by
Eq.(\ref{V3}), for the Henon map, after $t=2$ iterations, with an initial perturbation along the $x$ direction, $\delta \un{x}_0=(2.5\cdot 10^{-2},0)$, and different numbers of bins $Z_x$ on the $x$-axis, as well as different number $N$ of trajectories, so to 
keep the statistics in the $\epsilon \rightarrow 0$ limit: $N=3\cdot10^6, ~ Z_x=1320$ (green curve), 
$N=5\cdot 10^6, ~ Z_x=1650$ (purple curve), $N=8\cdot 10^6, ~ Z_x=1980$ (blue curve). \textit{Right panel}: plot of the function $B_y^{(q)}(y,\delta \un{x}_0,t,\epsilon)$ after $t=2$ iterations, for the same initial perturbation considered in the left panel, and different values of bins $Z_y$ on the $y$-axis and different values of $N$: $N=3\cdot10^6,~  Z_y=336$ (green curve), $N=5\cdot 10^6, ~ Z_y=420$ (purple curve), $N=8\cdot 10^6, ~ Z_y=505$ 
(blue curve).  The curves largely overlap, but 
the figure does not clarify whether $B_x^{(q)} , ~ B_y^{(q)}$ 
get smoother as $Z_x$ and $Z_y$ increase with $N$.}
\label{Henon}
\end{figure}
\begin{figure}
   \centering
   \includegraphics[width=8.5cm]{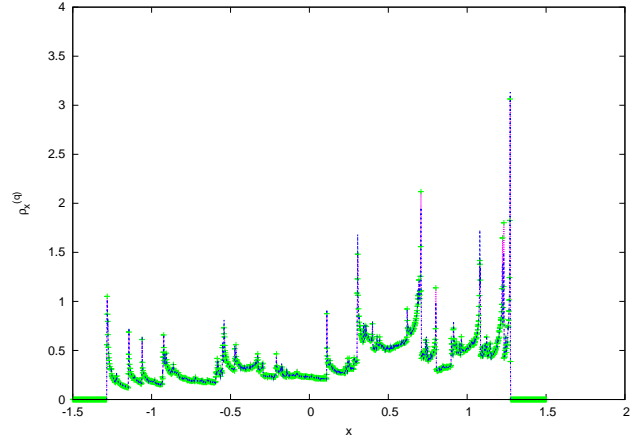}
   \caption{{Plot of the projected invariant probability density $\rho_x^{(q)}(\epsilon)$, for the Henon map,
with $N=5\cdot10^6$, $Z_x=1650$ (purple curve, with green errorbars) and  
$N=8\cdot10^6$, $Z_x=1980$ (blue curve). The figure does not clarify whether the invariant 
distribution is siungular or not.}}
\label{Henonerr}
\end{figure}

Then, because no direction appears to be priviledged in phase space, an initial perturbation 
along one of the axis should not lead to any singular perturbed projected measure, or
irregular response function, see e.g.\  Refs.\ \cite{EvRon,BKL2005}. Unfortunately, this 
is not obvious from the histograms constructed with growing numbers of bins, as they seem to
be quite irregular and to develop singularities in some parts of the phase space, 
cf.\ Figs.\ \ref{Henon} and \ref{Henonerr}. However, this does not necessarily prevent
the projected measures from having a density.

\begin{figure}
\centering
\includegraphics[width=10.5cm]{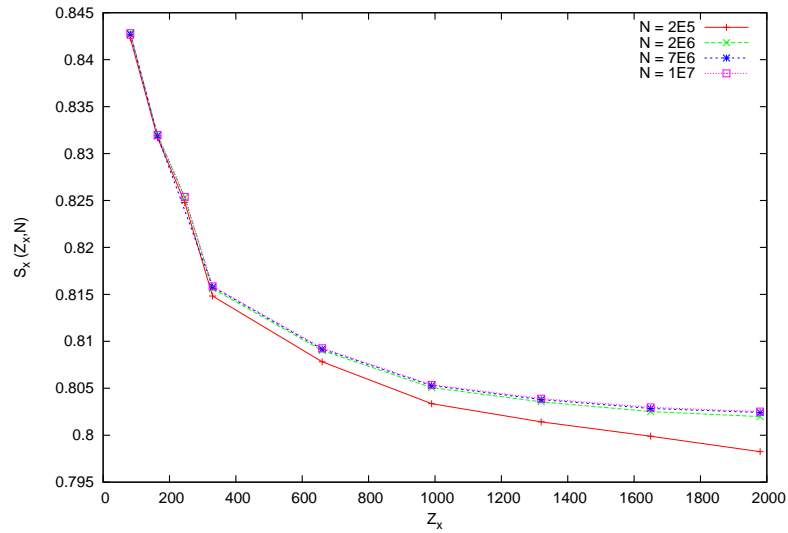}
\caption{Numerical simulations of the Shannon Entropy $S_x(\epsilon)$ for the Henon map, where $Z_x$ denotes the number of bins considered on the $x$-axis and $N$ denotes the number of trajectories. The curves collapse onto each other, approximating the asymptotic value of $S_x$ associated with 
the $N,Z_x\rightarrow \infty$ limits.}\label{Entropy}
\end{figure}

\begin{figure}
\centering
\includegraphics[width=6.5cm]{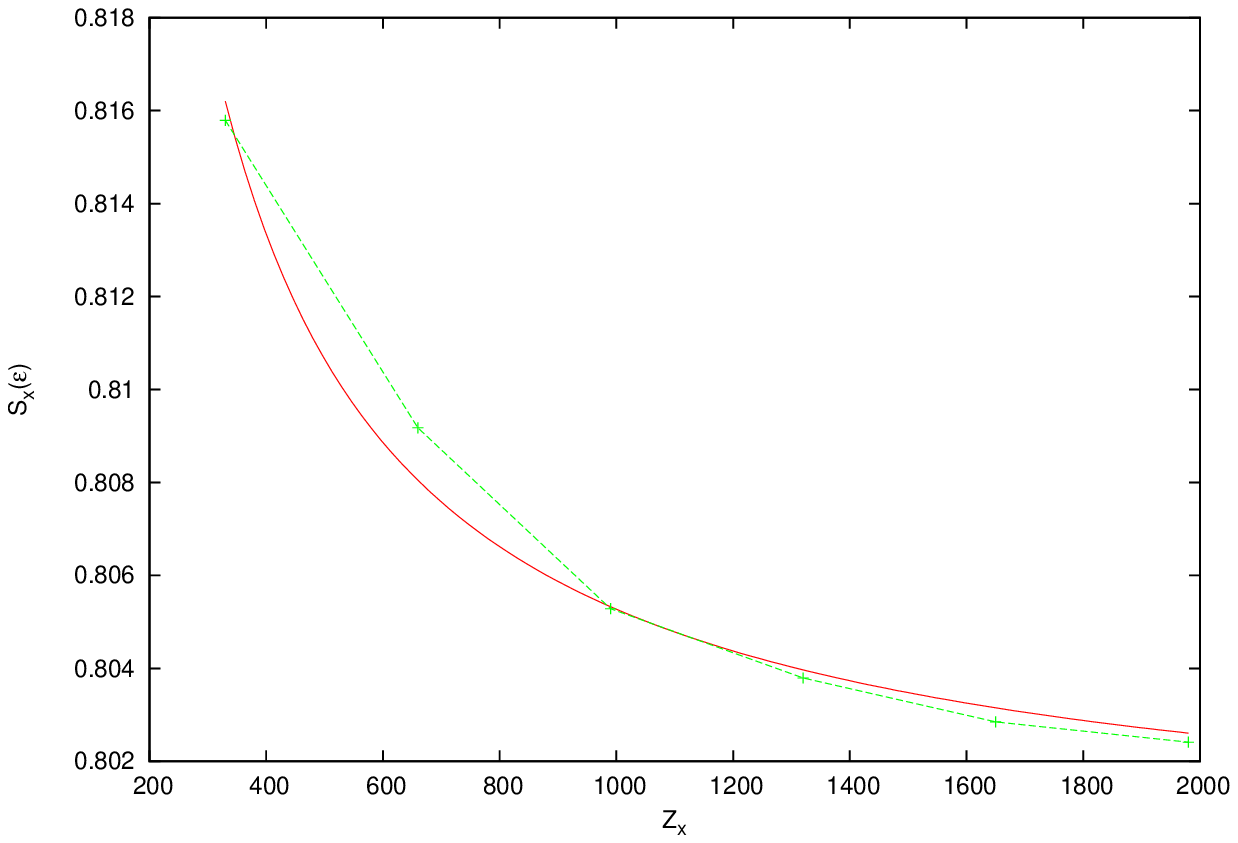}
\hspace{2mm}
\includegraphics[width=6.5cm]{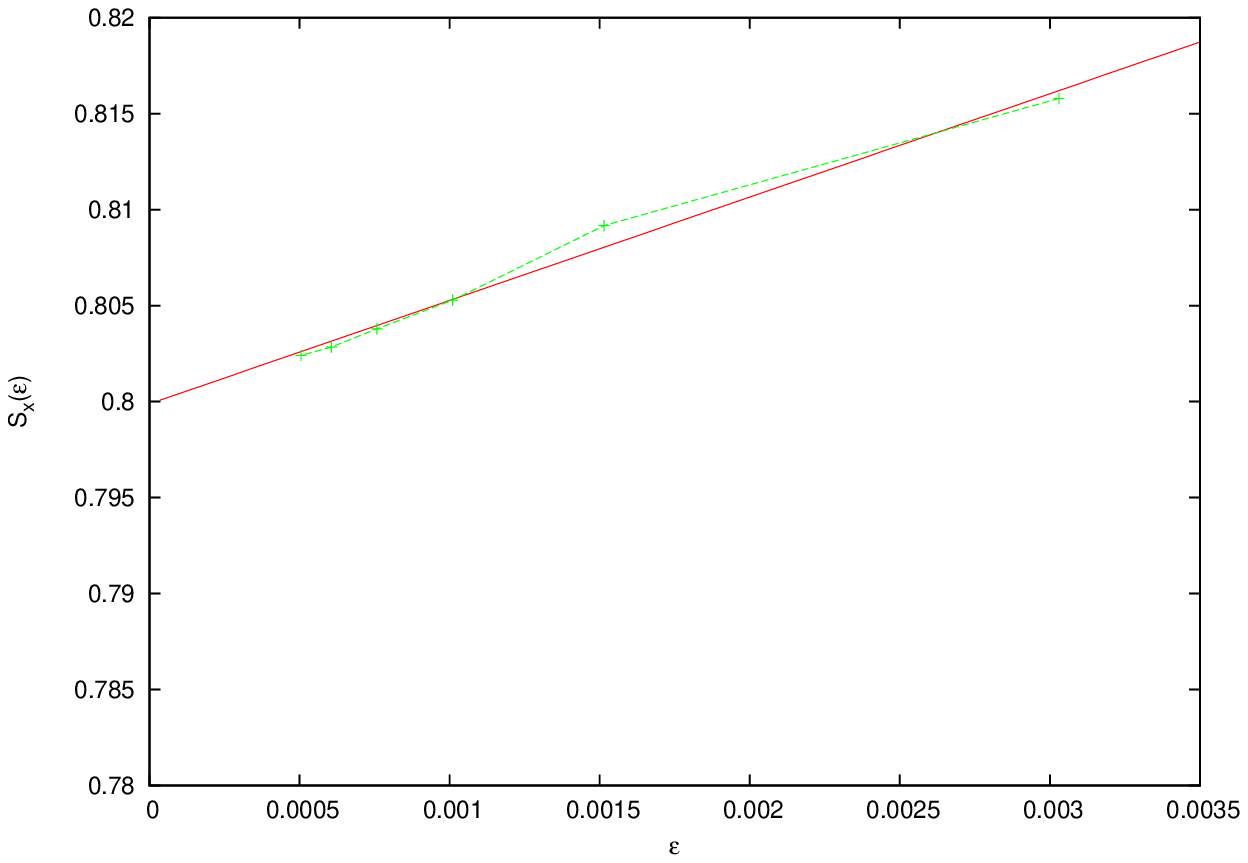}
\caption{\textit{Left panel}: Fit of the numerical data for $S_x$ corresponding to $N =7\cdot 10^6$, with $f(x)=a_0+\frac{a_1}{Z_x}$, $a_0=0.7998$ and $a_1=5.3847$. 
\textit{Right panel}: Same curve as in the left panel, plotted vs.\ the variable $\epsilon= 1/Z_x$, to extrapolate the asymptotic value $S_x \simeq 0.8$.}
\label{Entropyfit}
\end{figure}

Therefore, to clarify whether the projected probability density $\rho_x^{(q)}(\epsilon)$ 
exists or not in the $\epsilon \rightarrow 0$ limit, we have examined the behavior 
of the Shannon Entropy, defined as
\be
S_i(\epsilon)=-\epsilon\sum_{q=1}^{Z_i}\rho_i^{(q)}(\epsilon)\log(\rho_i^{(q)}(\epsilon)) \label{Shan}
\ee
with $\epsilon$ the size of the bin along the direction of the perturbation. Note that this entropy
is often defined differently; our definition is meant to introduce a quantity whose $\epsilon\rightarrow 0$ 
limit is finite if a density exists, while it diverges if the measure is singular.
We approximated $S_x(\epsilon)$ by running different sets of trajectories, with different sizes of the coarse graining of the $x$-axis. Our simulations with $N=2\cdot 10^6$, show that $S_x$ has substantially converged to its asymptotic $N\rightarrow \infty$ limit, cf. Fig.\ \ref{Entropy}.
Moreover, for fixed $N$, $S_x$ decreases as the number $Z_x$ of bins grows, and appears
to tend to a constant as $1/Z_x\rightarrow 0$, cf.\ Fig.\ \ref{Entropyfit}.

\begin{figure}
\centering
\includegraphics[width=6cm]{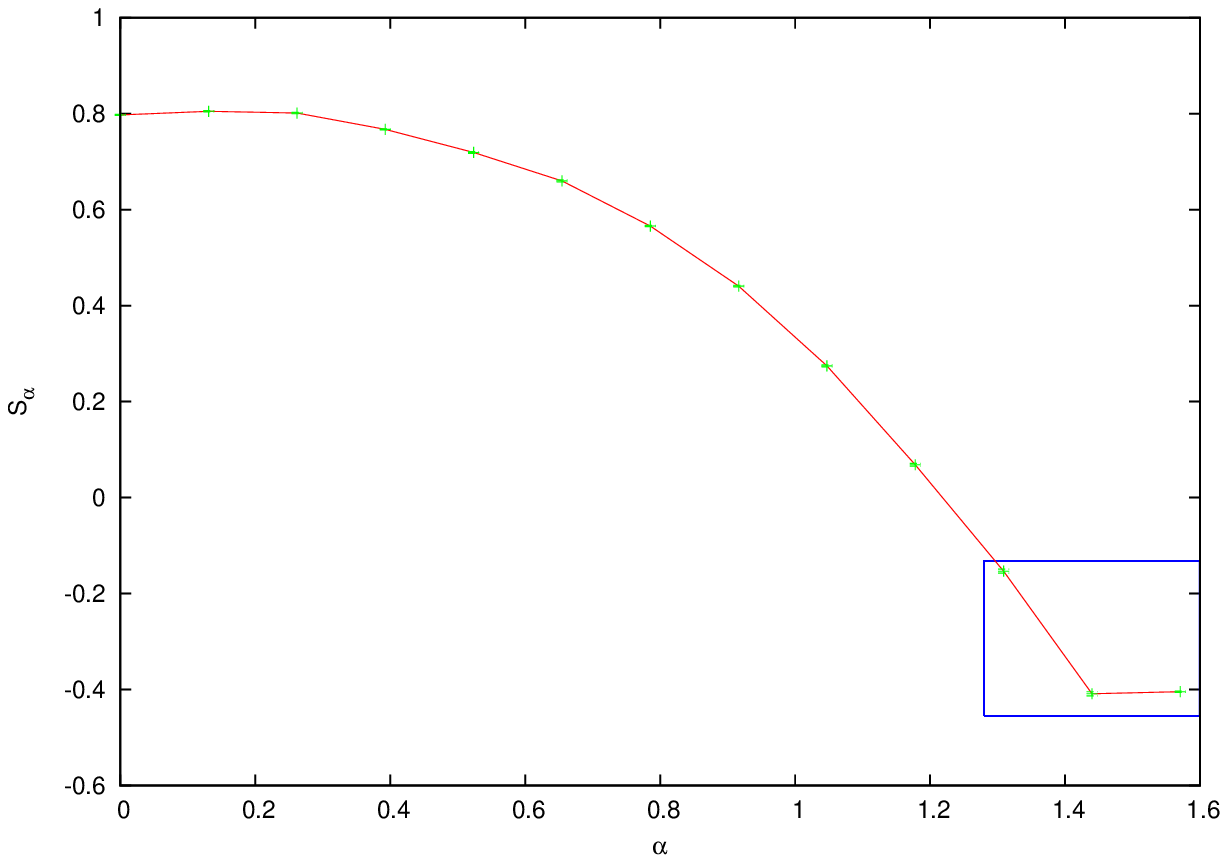}
\hspace{2mm}
\includegraphics[width=6cm]{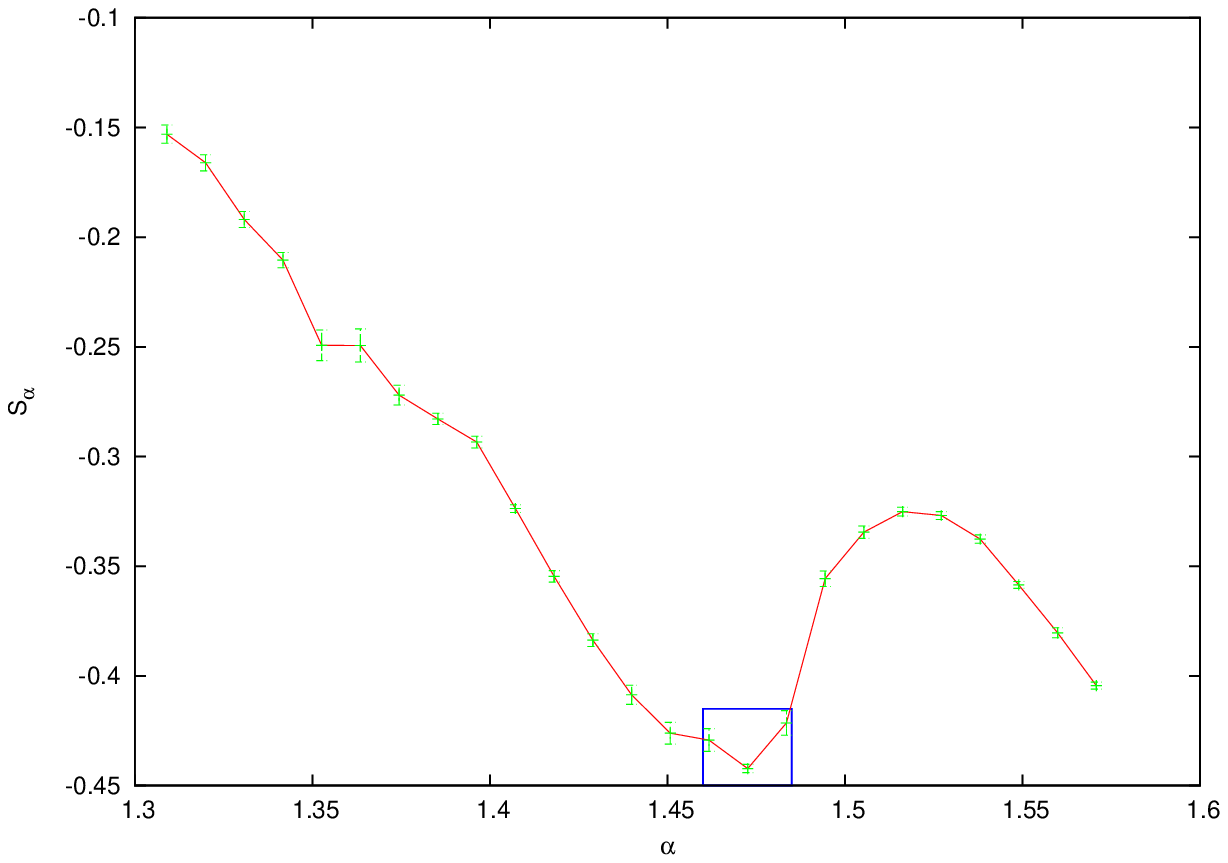}
\hspace{2mm}
\includegraphics[width=6cm]{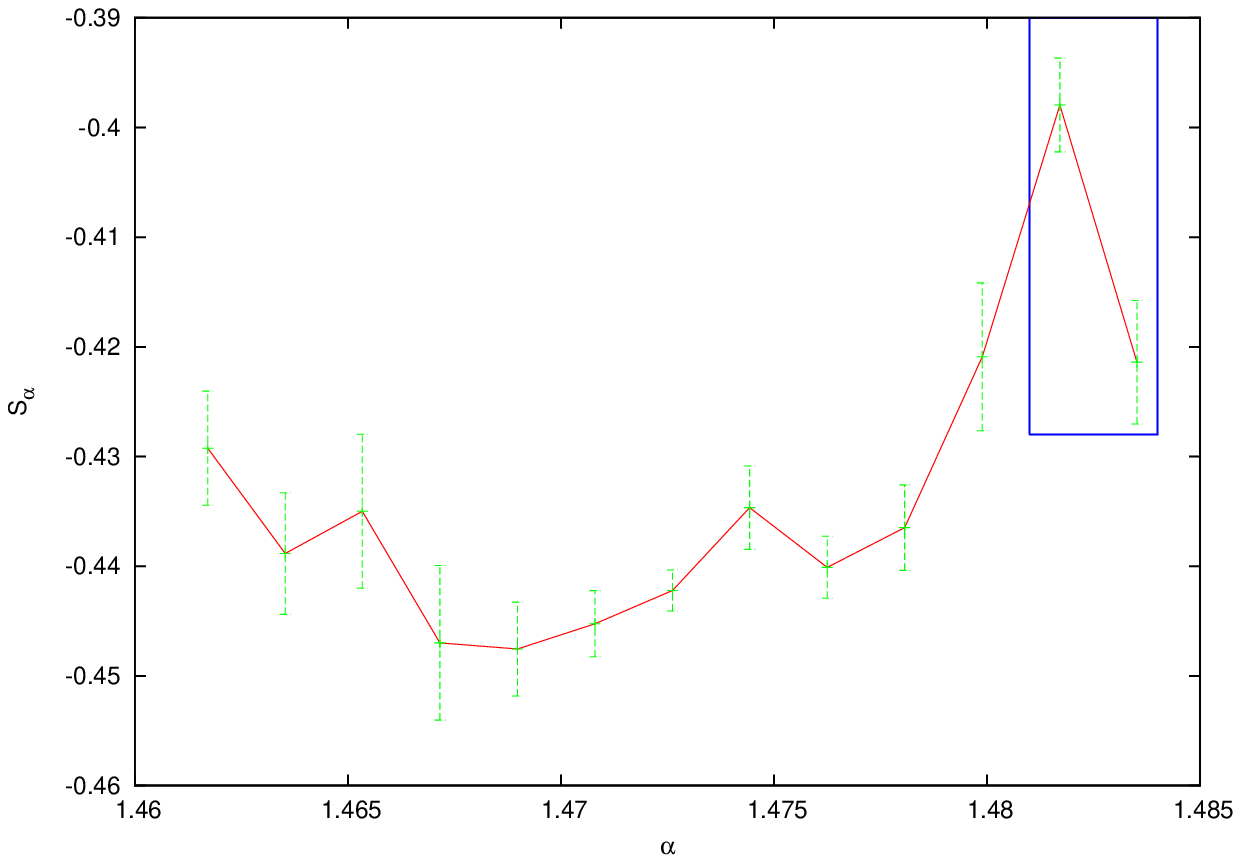}
\hspace{2mm}
\includegraphics[width=6cm]{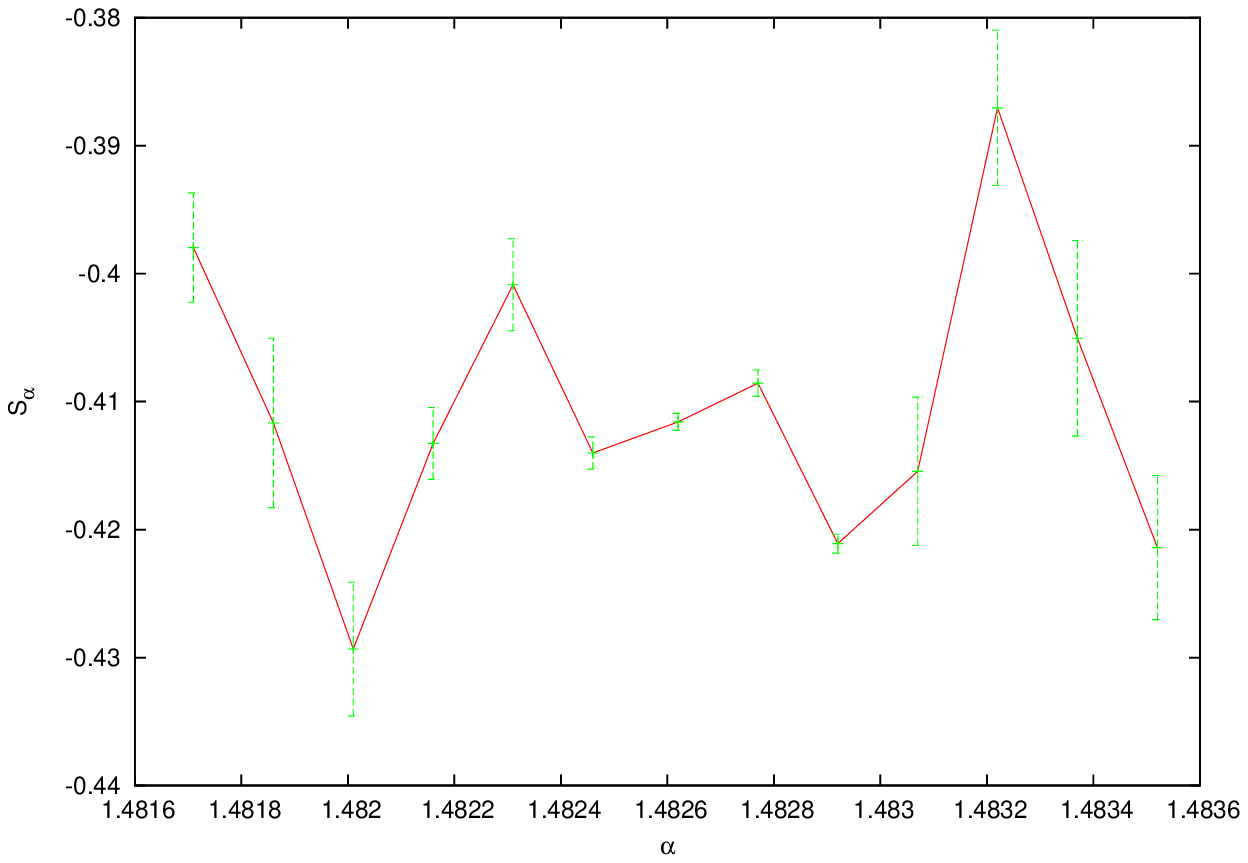}
\caption{Numerical data with error bars, for the Shannon Entropy $S_{\alpha}$ of the 
Henon map (\ref{Map2}), as a function of the angle $\alpha\in[0,\pi/2]$ with respect to
the horizontal direction, at increasing orders of magnification. The top right panel is 
a magnification of the framed part of the top left panel. The bottom left panel is a magnification of 
the framed part of the top right panel. The bottom right panel is a magnification of the 
framed part of the bottom left panel. The quantity $S_{\alpha}$ is quite structured, especially for $\alpha$ close to $\pi/2$, but 
it is a finite quantity.}
\label{SalHenon}
\end{figure}

\begin{figure}
\centering
\includegraphics[width=7.2cm]{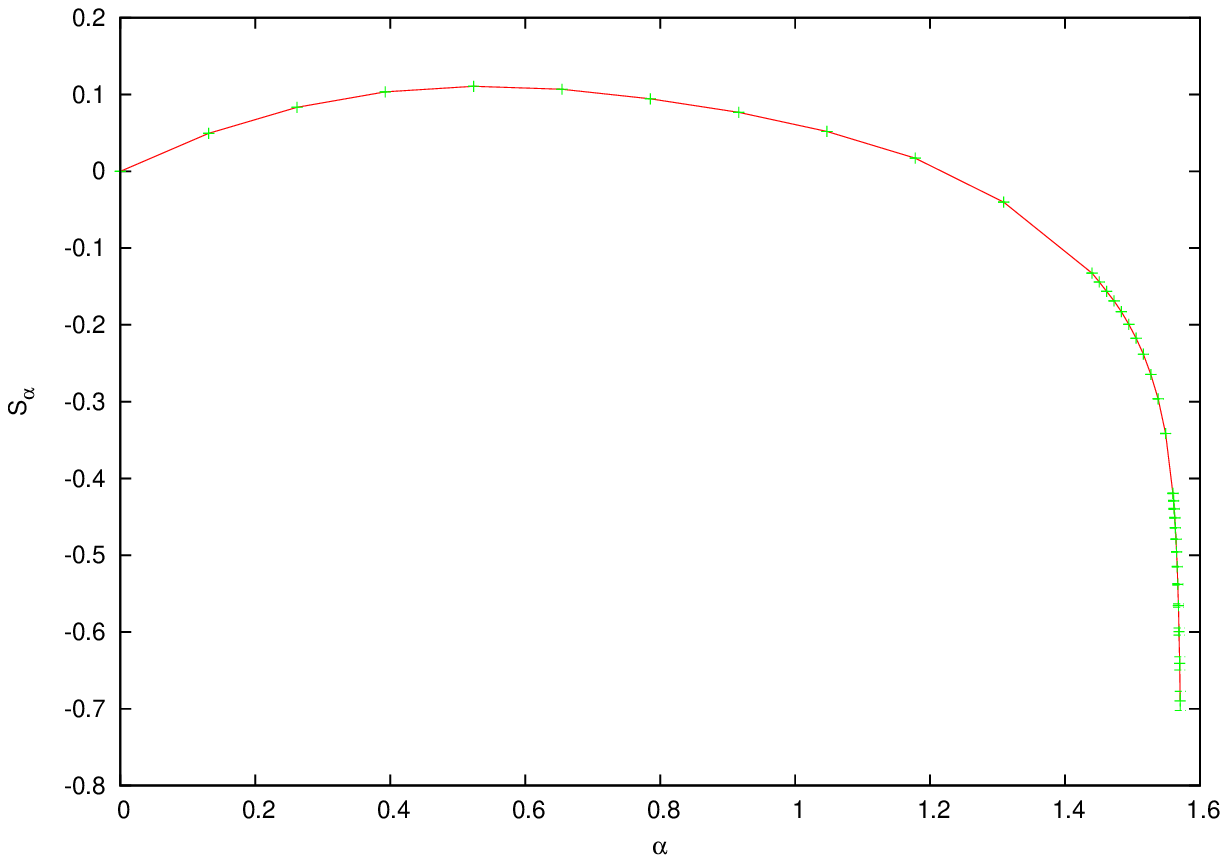}
\hspace{2mm}
\includegraphics[width=7.2cm]{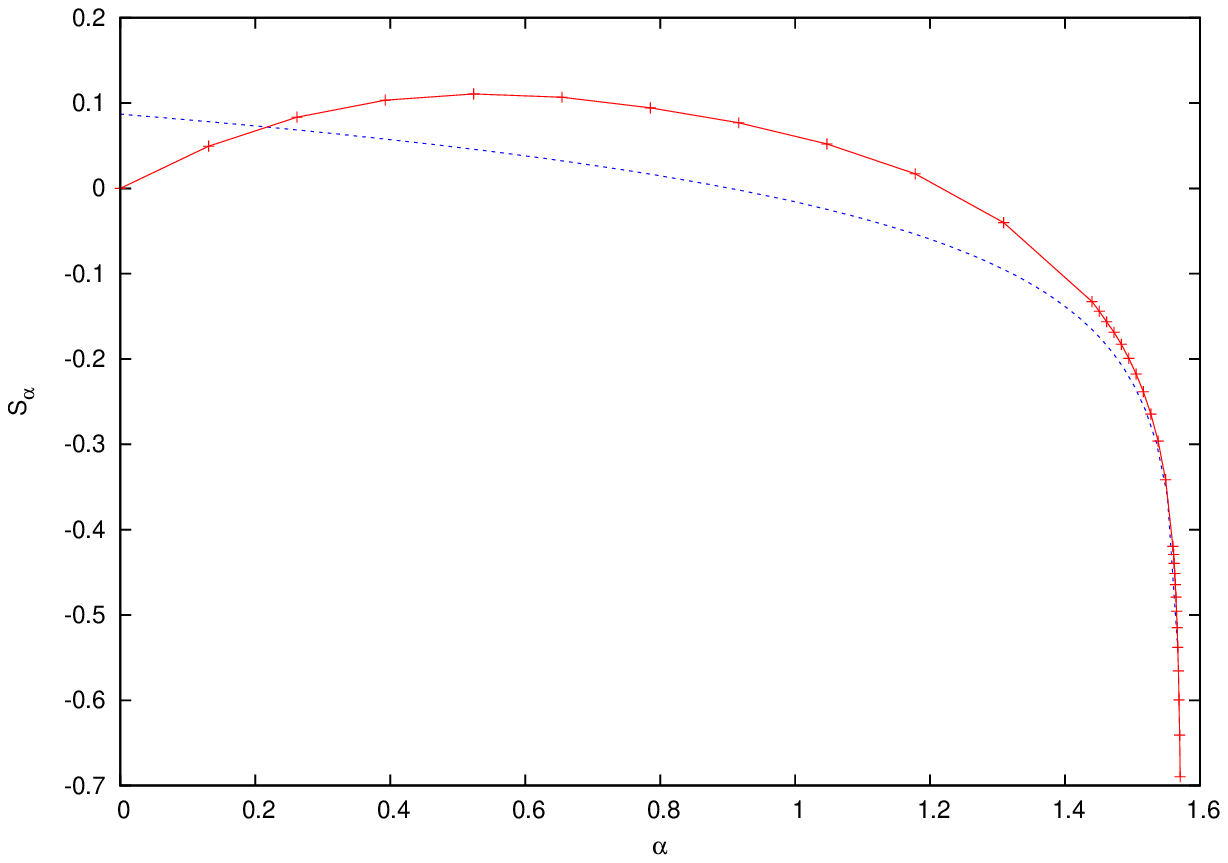}
\caption{\textit{Left panel}: Numerical data, supplemented by error bars, for the Shannon Entropy
$S_{\alpha}$ of the Baker map (\ref{Map1}), as a function of the angle $\alpha\in[0,\pi/2]$ with respect to the $x$-axis. 
\textit{Right panel}: Fit of $S_{\alpha}$ for the Baker map, with the curve 
$f(\alpha)=a \log(\frac{\pi}{2}-\alpha)+b$, $a=-0.101524\pm0.002768$, and $b=0.0411212\pm0.01401$.
In this case, $S_\alpha$ is a much simpler function of $\alpha$ than in the Henon case, and it
diverges in the $\alpha \to \pi/2$ limit, because the projected invariant measure is singular at 
$\alpha=\pi/2$.}
\label{SalBaker}
\end{figure}

Figures \ref{SalHenon} and \ref{SalBaker} further prove that $S_\alpha$ is always a finite 
quantity in the Henon case while, in the baker case, it diverges logarithmically only when 
$S_\alpha$ tends to $\pi/2$,
which is the only angle for which the projected invariant measure is singular. Therefore, 
the response can be obtained from the invariant measure at all perturbation angles in 
the case of the Henon map, and at all but a single angle for the Baker map.
This confirms the applicability of a generalized FDT, which yields the response function in 
terms of the unperturbed state only, even if supported on a fractal set, except in very
special situations, such as a negligible set in cases in which the invariant measure is
the product of regular and singular mesaures.
In particular, for the baker map, the response to a perturbation may be expressed just in terms 
of the smooth projected invariant measure if one does not perturb uniquely the vertical coordinates.
For the Henon map, all directions lead to the existence of a projected invariant measure,
although very finely structured.

\section{Conclusions}
\label{sec:sec4}
In this paper we have reviewed the methods proposed by Ref.\ \cite{Ruelle} and by Refs.\ 
\cite{BPRV,BLMV}, concerning the derivation of response formulae for systems 
in nonequilibrium steady states. In particular, we have shown that the idea of \cite{BLMV}, 
which is based on the existence of a smooth invariant probability density, may be applied quite
generally, with suitable adjustments, to dissipative deterministic dynamics. This requires
that projected distributions be considered, rather 
than the full phase space distributions, becuase projected distributions are usually regular. 
Only very special combinations of
dynamics and perturbations seem to prevent this approach, although in low dimensional 
dynamics such as ours, the projected distribution functions appear to be quite complex 
and not smooth.

The presence of noise, in any physically relevant dynamical system, does contribute to 
smooth out the invariant density, but even in the absence of noise, the fact that statistical 
mechanics is typically interested in projected dynamics allows an approach to FDT which only 
requires the properties of the unperturbed states, as in standard response theory. 
Clearly, this is better and better justified as the dimensionality of the phase space grows. 
In particular, it is appropriate for macroscopic systems in nonequilibrium steady states, because the
dynamics of interest take place in a space whose dimensionality is enormously smaller
than that of the phase space. Then, as projecting out more and more produces smoother 
and smoother distributions, one finds that the approach of Ref.\cite{BLMV} 
can be used to obtain the linear response function about nonequilirium steady states, 
from the unperturbed measure only.

Our results support the idea that the projection procedure makes unnecessary the explicit 
calculation of the term discovered by Ruelle, which was supposed to forbid the standard 
approach. This does not mean that Ruelle's term is necessarily negligible \cite{Sepul}. 
However, except in very peculiar situations, such as our maker map which has carefully 
oriented manifolds, and for carefully chosen perturbations, that term does 
not need to be explicitly computed and the calculation of response may be carried out referring 
only to the unperturbed dynamics, as in the standard cases.

\end{document}